\documentclass[aps,twocolumn,nofootinbib]{revtex4-1}
\usepackage{graphicx}
\usepackage{hyperref}
\usepackage{amsmath}
\usepackage{amssymb}

\newcommand{\beq}{\begin{eqnarray}}
\newcommand{\eeq}{\end{eqnarray}}
\def\simlt{\stackrel{<}{{}_\sim}}

\begin{document}

\title{A Dark-Disk Universe}

\author{{JiJi Fan, Andrey Katz, Lisa Randall, and Matthew Reece}}
\affiliation{Department of Physics, Harvard University, Cambridge, MA, 02138}

\begin{abstract}
 We point out that current constraints on dark matter imply only that the majority of dark matter is cold
and collisionless. A subdominant fraction of dark matter could have much stronger interactions. In particular,
it could interact in a manner that dissipates energy, thereby cooling into a rotationally-supported disk, much as baryons do.
We call this proposed new  dark matter component  Double-Disk Dark Matter (DDDM). We argue that DDDM could constitute a
fraction of all matter roughly as large as the fraction in baryons, and that it could be detected through its gravitational effects on
the motion of stars in galaxies, for example. Furthermore, if DDDM can annihilate to gamma rays, it would give rise to an indirect detection
signal distributed across the sky that differs dramatically from that predicted for ordinary dark matter.
DDDM and more general partially interacting dark matter scenarios provide a large unexplored space of testable new physics ideas.
\end{abstract}

\maketitle


{\bf Introduction.}
  Most of the matter in the universe is dark, distributed in diffuse halos around galaxies. Even so, the subdominant component consisting of baryons, electrons,
 and photons---the stuff of everyday life---though constituting only about 5\% of the universe's energy density,
gives rise to rich phenomena in the world around us. Our goal in this paper is to argue that dark matter too could contain a component exhibiting
diverse and observable consequences: the dark world
might even be  as diverse
and interesting as the visible world. This hypothesis is worth exploring as it can be tested in several complementary ways.

 The structure of our galaxy relies on interacting baryons that
can cool. They do so by dissipating energy through photon emission as they collapse to form structure.
Cooling is a prerequisite to baryonic structures occupying relatively
small volumes and forming compact objects like stars and planets. On a larger scale, it is
 necessary for the formation of disk galaxies.

In stark contrast to baryons, we typically
assume that dark matter (DM) is cold and collisionless, distributed through a large halo in a random way.
This paradigm is sometimes relaxed as in the cases of self-interacting dark matter (SIDM)~\cite{Spergel:1999mh}
 or warm dark matter~\cite{Bode:2000gq}, but such scenarios are bounded by observations of halo shapes and the Bullet
 Cluster that limit the amount by which dark matter can deviate from being cold and collisionless. These bounds are often
thought to imply that the world of dark matter is much less rich and interesting than the world of visible matter, and as a
 result dark matter is usually assumed to be a single type of particle, like a WIMP.

In this paper we propose that the dark world could be as
complex as the visible world, with a simple assumption: while most
of the dark matter is cold and collisionless, a subdominant fraction we call
Partially Interacting Dark Matter (PIDM) could interact more strongly and even cool as baryons do.
 This subdominant fraction could have an energy density about as large as that of baryons, without having
been noticed so far. If its dynamics are dissipative, it will cool and form a disk within galaxies, much as baryons do.
Our own Milky Way could contain structures made of interacting dark matter, analogous to the structures in the visible world around us,
in an invisible disk parallel to our own. We call this possibility Double-Disk Dark Matter (DDDM). In this letter, we  outline
the physics of DDDM and some of the observational possibilities. In a companion paper~\cite{OurLongerPaper}, we provide more detailed
 calculations and consistency checks of the scenario. Both DDDM and the more general idea of PIDM raise a large number
of interesting questions, which we have only begun to explore. 

Dark disks may also arise from ordinary DM accreting onto the stellar disk~\cite{Read:2008fh, Bruch:2008rx, Bruch:2009rp}. Their phenomenology of direct detection and solar capture are similar, but our mechanism to generate the disk is completely different.

\begin{figure*}[!t]\begin{center}
\includegraphics[width=1.7\columnwidth]{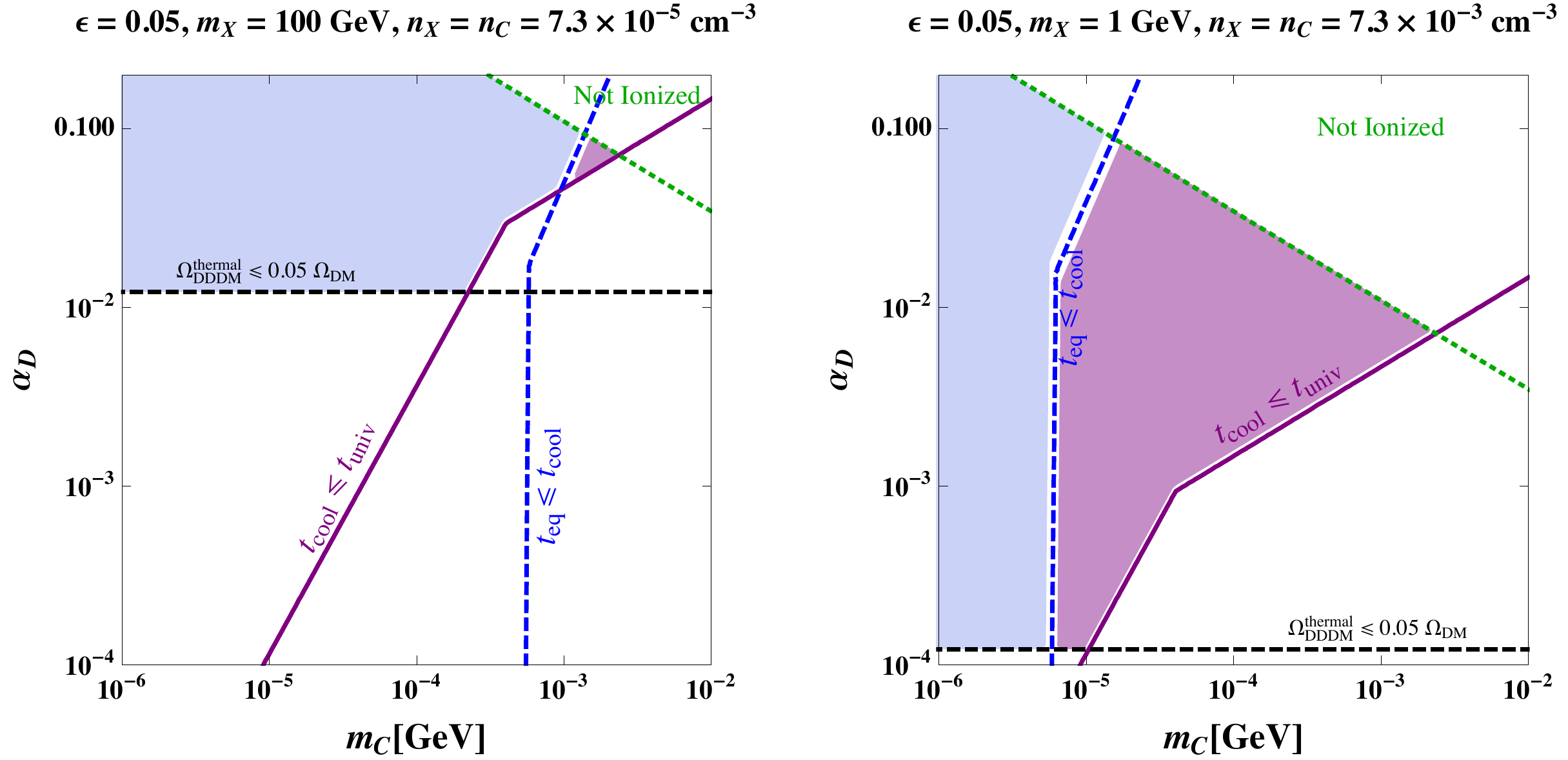}
\end{center}
\caption{Cooling in the $(m_C, \alpha_D)$ plane. The purple shaded region cools adiabatically within the age of the universe. The light blue region cools out of equilibrium. We take redshift $z = 2$ and $T_D = T_{\rm CMB}/2$. At left, $m_X = 100$ GeV; at right, $m_X = 1$ GeV. The density chosen corresponds to a 20 kpc NFW virial cluster. The solid purple curves show where the cooling time equals the age of the universe; they have a kink where Compton-dominated cooling (lower left) transitions to bremsstrahlung-dominated cooling (upper right). The dashed blue curve delineates fast equipartition of heavy and light particles. Below the dashed black curve, small $\alpha_D$ leads to a thermal relic $X,{\bar X}$ density in excess of the Oort limit. To the upper right of the dashed green curve, the $XC$ binding energy is high enough that dark atoms are not ionized and cooling would be through atomic processes we do not calculate.}
\label{fig:rate}
\end{figure*}%


{\bf Bounding the Amount of DDDM.} In a scenario with both ordinary (cold, collisionless) dark matter and a
 more strongly interacting component, we denote the fraction in the interacting component by $\epsilon \equiv \Omega_{\rm PIDM}/\Omega_{\rm DM}$. In the case of DDDM, we denote the fraction of the Milky Way's mass localized in a dark disk by $\epsilon_{\rm disk} \equiv M^{\rm disk}_{\rm DDDM}/M^{\rm gal}_{\rm DM}$. Given that only about a third of baryons end up in the galactic disk, $\epsilon_{\rm disk}$ might be up to about three times less than $\epsilon$. Some baryons are removed from the baryonic disk due to feedback from supernovae~\cite{Dutton:2008hh}, absent from the DDDM sector. Model-dependent compact DDDM objects may lead to feedback and deserve future study, e.g. for how DDDM affects the Tully-Fisher relation, for which baryonic outflow can be important~\cite{Dutton:2008hh}.
 
 Current
bounds on self-interacting dark matter arise from halo shapes and cluster interactions and have been applied only when {\em all} the DM
is self-interacting, for which they can be quite constraining. Self-interactions give halos a spherical core
 and can be in tension with observations of elliptical halos~\cite{Dave:2000ar,Rocha:2012jg}.
However, such bounds do not directly apply to PIDM since a sufficiently small fraction of all matter could have
extremely strong interactions without affecting observations. Self-interactions are also bounded by the Bullet
Cluster~\cite{Markevitch:2003at}, which displays a separation between (collisional) hot gas and collisionless material
(stars and ordinary dark matter). The observations imply that no more than 30\% of the dark matter was lost to collisional effects,
i.e. $\epsilon \leq 0.3$.

A stronger bound arises when dissipation and hence cooling occurs with the consequent formation of a dark disk.  The total amount of matter in the neighborhood of the
 Sun is measured and known as the Oort limit. According to~\cite{Weber:2009pt}, the total surface density in the Milky Way
near the Sun, $\Sigma \equiv \int_{-1.1~{\rm kpc}}^{+1.1~{\rm kpc}} \rho(z) dz$, is measured as $71 \pm 6~M_\odot/{\rm pc}^2$.
The surface density accounted for in visible matter is smaller, between 35 and 58$~M_\odot/{\rm pc}^2$. Comparing these two numbers,
 we find that a surface density in dark matter as large as $46~M_\odot/{\rm pc}^2$ is allowed by the data at 95\% CL. We model the DDDM
disk as an isothermal sheet~\cite{MvdBW}:
\beq
\rho(R,z) = \frac{\epsilon_{\rm disk} M^{\rm gal}_{\rm DM}}{8\pi R_d^2 z_d} \exp(-R/R_d) {\rm sech}^2(z/2z_d).
\label{eq:distribution}
\eeq
When the disk height $z_d \ll 1.1~{\rm kpc}$, the surface density is $z_d$-independent: $\Sigma_{\rm disk} = \epsilon_{\rm disk} M^{\rm gal}_{\rm DM}/(2\pi R_d^2) \exp(-R/R_d)$. We take the scale radius of the disk to be similar to that for baryons, $R_d \approx 3$ kpc; then the bound on $\Sigma$ implies:
\beq
\epsilon_{\rm disk} \simlt 0.05.
\eeq
In other words, we estimate that as much as 5\% of the DM in the galaxy can be localized in a thin disk.
This matches the mass of the baryonic disk and implies the DDDM density in the universe can be comparable to that of baryons. Measurements of kinematics of visible objects, like the billion stars
the Gaia satellite~\cite{Famaey:2012ga} will measure, might detect the gravitational effects of such a structure.


{\bf Model and Early Cosmology.} We construct DDDM to mimic baryonic matter in many respects. The simplest such model has a heavy field $X$
and light field $C$ that are charged $+1$ and $-1$ respectively
 under a gauged U(1)$_D$ with coupling strength $\alpha_D$. Kinetic mixing bounds can be circumvented in several ways~\cite{OurLongerPaper}.
Nonabelian models with small coupling constant are also suitable.  We will
 typically take $m_C \sim 1$ MeV and $m_X \sim 1$ to 100 GeV. Related scenarios include Hidden Charged Dark Matter~\cite{Feng:2008mu} and Atomic
Dark Matter~\cite{Goldberg:1986nk,Kaplan:2009de,CyrRacine:2012fz}. Our innovation is considering that this sector may constitute only a fraction
of the dark matter and so can have dissipative dynamics without conflicting with data.\footnote{Other scenarios, like Dynamical Dark Matter~\cite{Dienes:2011ja}, are similar in spirit but differ in details.}

We assume that at early times the dark sector and the Standard Model were in thermodynamic equilibrium above the weak scale. As the universe cooled
 and Standard Model degrees of freedom decoupled from the thermal bath, the dark sector became cooler than the SM sector by a factor $\xi \equiv T_D/T_{\rm vis}$. Estimates
show that $\xi \approx 0.5$ at the times relevant for BBN and CMB observations. Bounds on relativistic degrees of freedom are typically expressed in
terms of effective neutrino species, with
95\% CL bounds $\Delta N_{{\rm eff}, \nu}^{\rm BBN } < 1.44$~\cite{Cyburt:2004yc} and $\Delta N_{{\rm eff},\nu}^{\rm CMB} < 1.0$~\cite{Ade:2013lta} (Planck+WP+highL+$H_0$+BAO).  
 In the case of a decoupling temperature between the $b$ and $W$ masses, we estimate $\Delta N_{{\rm eff},\nu} \approx 0.2$ for both BBN and the CMB,
increasing to $\approx 0.9$ in an SU(3)$_D$ model. At the time of the CMB, the dark photons and $C$ particles are coupled in a plasma, which can give rise to
 effects like dark acoustic oscillations~\cite{CyrRacine:2012fz}, which are left for future study. 

When the dark sector cools below about $m_C/20$, the $C$ and ${\bar C}$ particles  annihilate away, much as electrons and positrons did in our universe.
 As a result, we assume an asymmetric dark matter scenario~\cite{AsymmetricDM} with net $X$ and $C$ number, analogous to the proton and electron number in
our universe. For $\alpha_D < 0.01$, we find that a residual symmetric population of $X$ and ${\bar X}$ will not completely annihilate away at early times,
so in general we expect both asymmetric DM {\em and} a symmetric $X, {\bar X}$ component to remain. The latter is particularly interesting since it can provide an
indirect detection signal through annihilation processes like $X{\bar X}$ to gamma rays.


{\bf Cooling and Disk Formation.} Energy dissipation of $X$ and $C$ particles in galaxies is very similar to that of protons and electrons.
In the early universe, $X$ and $C$ may have bound into dark atoms, but when galaxies form and dark matter is shock-heated to the virial temperature,
which at large $m_X$ is even higher than that for baryons,
   a gas of ionized $X$ and $C$ particles distributed throughout the halo will remain. This dark plasma cools, primarily through
Compton scattering of $C$ particles on dark cosmic background photons and through dark bremsstrahlung, $XC \to XC\gamma_D$. The calculation of these
 cooling processes is as in ordinary QED~\cite{SpitzerJr.:1941}. Compton scattering dominates at early times, with a rate growing with redshift as $(1+z)^4$.
  In both cases,  the light $C$ particles  dominantly lose energy, at
a rate that is faster for smaller $m_C$.
In a large portion of parameter space, Rutherford scattering of $X$ and $C$ particles leads to a fast equipartition of energy, so that the cooling of $C$ particles also
affects $X$ particles and the entire dark plasma cools adiabatically. For very small $m_C/m_X$, the Rutherford scattering rate is slow enough that
cooling of $X$ particles  happens out of equilibrium, with more complicated dynamics we do not study here. The region of parameter space that cools efficiently
in an adiabatic way is shaded purple in Figure~\ref{fig:rate}, while the light blue shaded region cools out of equilibrium.

As with baryons, the plasma of $X$ and $C$ particles acquires angular momentum from tidal torques, so as it cools it forms a rotationally supported
disk. For the simplest DDDM models, details like star formation feedback are absent, but we expect the disk will form regardless~\cite{Vogelsberger:2011hs}
and that gravitational attraction
and initial conditions favor approximate alignment of the baryonic and dark disks. The scale height of the disk can be estimated from the Jeans equation for
an axisymmetric system:
\beq
z_d \approx \sqrt\frac{2 \overline{v_z^2}}{\pi G_N \rho_{\rm center}} \approx 1.2 \frac{\overline{v^2_z}}{10^{-6}} \frac{R_d}{\epsilon_{\rm disk}}.
\eeq
The vertical velocity dispersion $\overline{v_z^2}$ is set by the temperature at which cooling stops. We expect cooling to stop when bremsstrahlung and Compton cooling
 cease---namely when $X$ and $C$ particles are cold enough to bind into dark atoms, at temperatures $T_{\rm cooled} \sim 0.1 B_{XC}$ where $B_{XC} = \alpha_D m_C/2$
is the binding energy of a dark atom. This leads to an estimate:
\beq
z_d \approx 2.5~{\rm pc} \left(\frac{\alpha_D}{0.02}\right)^2 \frac{m_C}{1~{\rm MeV}} \frac{100~{\rm GeV}}{m_X}.
\label{eq:diskheight}
\eeq
For very small values of $\alpha_D$ and $m_C$, the dark CMB temperature $\sim 1 K$ could be larger than $0.1 B_{XC}$ and set the limit of possible cooling.
Heating effects or atomic and molecular cooling processes that we have not considered may also play a role. Nonetheless, eqn.~\ref{eq:diskheight} leads us
to expect that the disk of DDDM can be much thinner than the baryonic disk, though it can thicken due to collisional processes analogous to those for baryons.

\begin{figure}[!t]\begin{center}
\includegraphics[width=0.8\columnwidth]{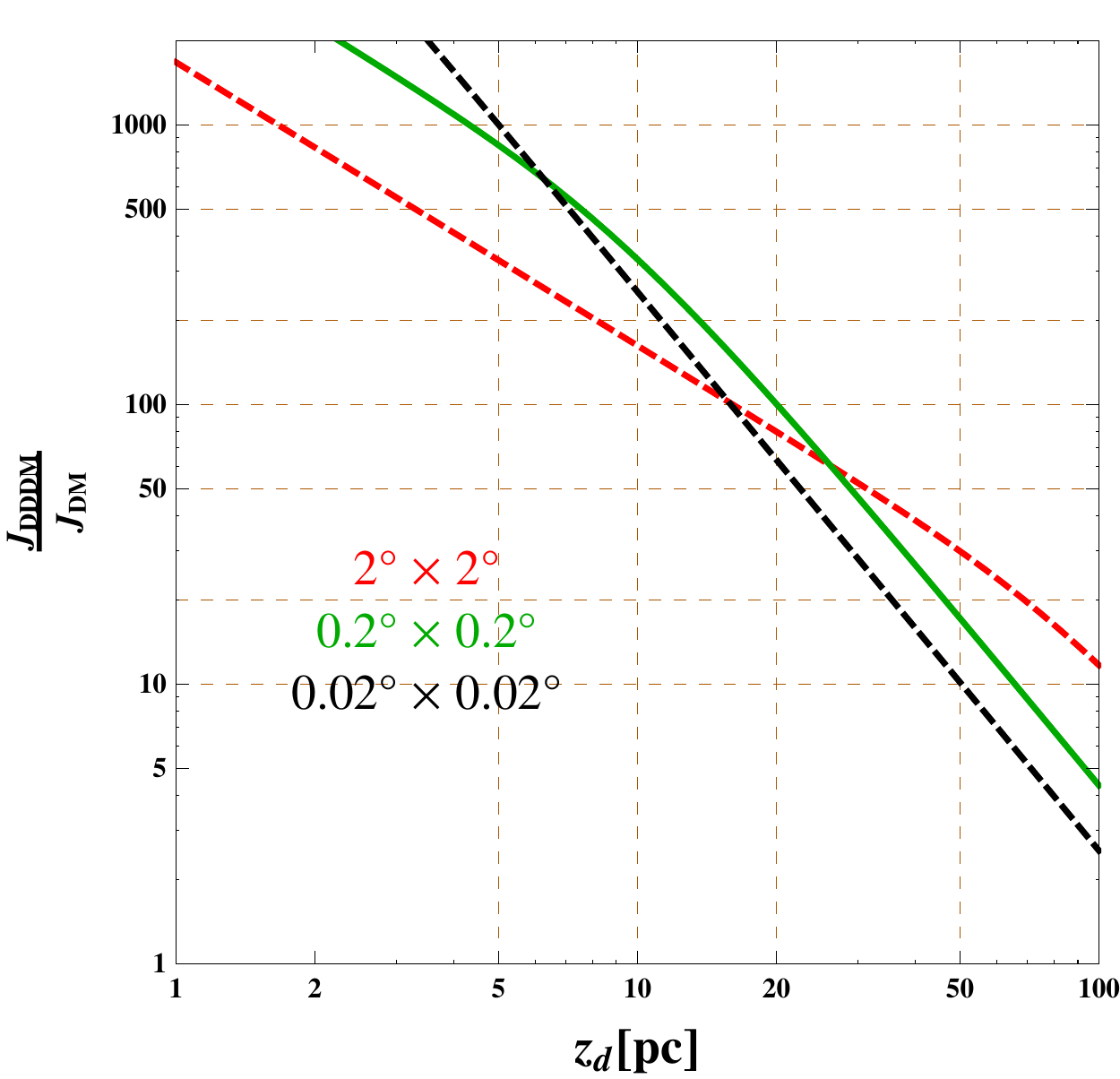}
\end{center}
\caption{Signal density enhancement $J_{\rm DDDM}/J_{\rm DM}$ for DDDM in a square region around the galactic center fixing $\epsilon_{\rm disk} = 0.05$. Red: region within $b,l \subset (-1^\circ, 1^\circ)$. Green: region within $b,l \subset (-0.1^\circ, 0.1^\circ)$ (current Fermi-LAT angular resolution). Black: region within $b,l \subset (-0.01^\circ, 0.01^\circ)$.  }
\label{fig:indirect}
\end{figure}%

\begin{figure}[!h]\begin{center}
\includegraphics[width=\columnwidth]{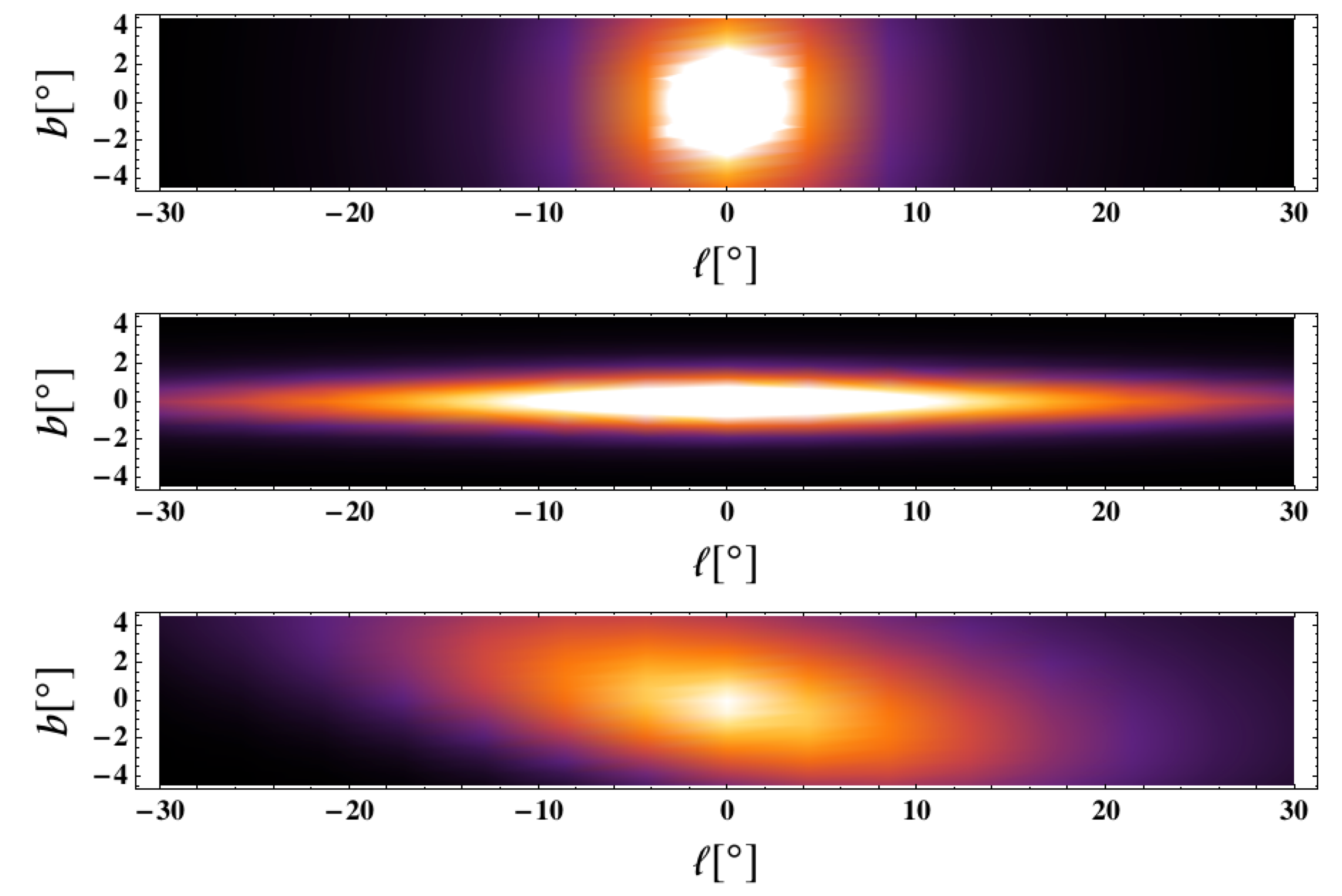}
\end{center}
\caption{Sky maps of the photon flux shape, in arbitrary units, for different DM profiles. Upper: Ordinary DM with an Einasto distribution.
 Middle: DDDM in a disk with scale height $z_d = 100$ pc, aligned with our disk. Lower:  DDDM in a similar disk, now misaligned with our disk
by 18$^\circ$.}
\label{fig:skymaps}
\end{figure}

{\bf Detection Prospects.} A thin dark disk would lead to a significant local density enhancement of DDDM compared to ordinary dark matter in
the plane of the galaxy. This may be detected mostly through gravitational effects if DDDM is all bound into dark atoms. But if there is a relic symmetric
population of $X$ and ${\bar X}$, as is expected for a large portion of parameter space, annihilation can produce indirect detection signals that are strikingly
 different from those of ordinary
dark matter. In the galactic center, there can be a significant enhancement of the line-of-sight integral of the dark matter number density squared,
\beq
J \equiv \int_{\rm roi} db \, dl \int_{\rm l.o.s} \frac{ds}{d_\odot}\cos b \left(\frac{\rho(r)}{\rho_\odot}\right)^2,
 \eeq
illustrated in Fig.~\ref{fig:indirect}. Such a large enhancement could provide the ``boost factor'' that is often proposed to make sense of the
size of tentative indirect detection signals such as from Fermi~\cite{Weniger:2012tx}, PAMELA~\cite{Adriani:2008zr}, or possibly AMS~\cite{Palmonari:2011zz}. The spatial distribution is even more distinctive, being extended across the sky due to the shape
of the disk, as shown in Fig.~\ref{fig:skymaps}.

In contrast to indirect detection, the prospects for direct detection suffer from two effects: first, the Earth might lie outside the
dark disk (due either to misalignment of the dark and baryonic disks, or the thinness of the dark disk). Second, even if the disks are aligned,
they will tend to be moving with the same average circular velocity, so the relative velocity of dark matter and nuclei will be low and the
DM will not have enough kinetic energy to exceed current experimental thresholds~\cite{Bruch:2008rx}. On the other hand, the slow relative velocity can enhance capture
in the Sun, so with improved neutrino observations detection might ultimately be possible~\cite{Bruch:2009rp}.


{\bf Conclusions.} DDDM is a rich and previously overlooked scenario for dark matter. DDDM in particular leads to rich and observable consequences. It will also be interesting to study chemistry or even more complex models with nuclear physics that would even more closely resemble ordinary matter with the creation of a shadow galactic disk.
 New results from Planck, the Gaia survey, and dark matter simulations will all contribute to a
 richer understanding of the structure of our universe, in ways previously unexplored.

{\bf Acknowledgments.} We would especially like to thank L. Hernquist for guidance early on in this project. We also thank A. Brown, C. Cheung, R. de Putter, D. Eisenstein, D. Finkbeiner, A.L. Fitzpatrick, J. Frieman, S. Genel, L. Hall, J. Kaplan, M. Kaplinghat, J. March-Russell, P. Mauskopf, M. McQuinn, M. Milgrom, A. Nelson, Y. Nomura, A. Nusser, J. Ruderman, M. Schwartz, T. Vachaspati, M. Walker, and R. Windhorst for useful discussions. We thank H. Georgi for naming DDDM. We are supported in part by the Fundamental Laws Initiative of the Harvard Center for the Fundamental Laws of Nature. AK and MR thank the GGI in Florence, Italy, for its hospitality while a portion of this work was completed. The work of LR was supported in part by NSF grants PHY-0855591 and PHY-1216270.

\end{document}